\documentclass[sigconf]{acmart}

\copyrightyear{2026}
\acmYear{2026}
\setcopyright{cc}
\setcctype{by-nc-nd}
\acmConference[SIGIR '26] {Proceedings of the 49th International ACM SIGIR Conference on Research and Development in Information Retrieval}{July 20--24, 2026}{Melbourne, VIC, Australia.}
\acmBooktitle{Proceedings of the 49th International ACM SIGIR Conference on Research and Development in Information Retrieval (SIGIR '26), July 20--24, 2026, Melbourne, VIC, Australia}
\acmISBN{979-8-4007-2599-9/2026/07}
\acmDOI{10.1145/3805712.3808436}

\begin{document}

\title{SID-Coord: Coordinating Semantic IDs for ID-based Ranking in Short-Video Search}

\author{Guowen Li}
\orcid{0009-0004-6597-7098}
\affiliation{
  \institution{Kuaishou Technology}
  \city{Beijing}
  \country{China}
}
\email{liguowen@kuaishou.com}

\author{Yuepeng Zhang}
\orcid{0009-0000-4923-2644}
\affiliation{
  \institution{Kuaishou Technology}
  \city{Beijing}
  \country{China}
}
\email{zhangyuepeng03@kuaishou.com}

\author{Shunyu Zhang}
\orcid{0009-0000-1936-1162}
\affiliation{
  \institution{Kuaishou Technology}
  \city{Beijing}
  \country{China}
}
\email{zhangshunyu@kuaishou.com}
\authornote{Corresponding authors.}

\author{Yi Zhang}
\orcid{0009-0009-0975-6369}
\affiliation{
  \institution{Kuaishou Technology}
  \city{Beijing}
  \country{China}
}
\email{zhangyi49@kuaishou.com}

\author{Xiaoze Jiang}
\orcid{0000-0002-5463-7176}
\affiliation{
  \institution{Kuaishou Technology}
  \city{Beijing}
  \country{China}
}
\email{xzjiang@buaa.edu.cn}

\author{Yi Wang}
\orcid{0009-0008-6816-9888}
\affiliation{
  \institution{Kuaishou Technology}
  \city{Beijing}
  \country{China}
}
\email{wangyi05@kuaishou.com}

\author{Jingwei Zhuo}
\orcid{0000-0001-8135-1061}
\affiliation{
  \institution{unaffiliated}
  \city{Beijing}
  \country{China}
}
\email{zhuojw10@gmail.com}
\authornotemark[1]
\renewcommand{\shortauthors}{Guowen Li et al.}

\begin{abstract}
    Large-scale short-video search ranking models are typically trained on sparse co-occurrence signals over hashed item identifiers (HIDs). While effective at memorizing frequent interactions, such ID-based models struggle to generalize to long-tailed items with limited exposure. This memorization–generalization trade-off remains a longstanding challenge in such industrial systems. We propose SID-Coord, a lightweight Semantic ID framework that incorporates discrete, trainable semantic IDs (SIDs) directly into ID-based ranking models. Instead of treating semantic signals as auxiliary dense features, SID-Coord represents semantics as structured identifiers and coordinates HID-based memorization with SID-based generalization within a unified modeling framework. To enable effective coordination, SID-Coord introduces three components: (1) an attention-based fusion module over hierarchical SIDs to capture multi-level semantics, (2) a target-aware HID–SID gating mechanism that adaptively balances memorization and generalization, and (3) a SID-driven interest alignment module that models the semantic similarity distribution between target items and user histories. SID-Coord can be integrated into existing production ranking systems without modifying the backbone model. Online A/B experiments in a real-world production environment show statistically significant improvements, with a +0.664\% gain in long-play rate in search and a +0.369\% increase in search playback duration.
\end{abstract}




\maketitle

\section{Introduction}
    In industrial short-video search ranking systems, sparse ID-based models trained on large-scale user–item interactions remain the dominant solution~\cite{QIN}. These models are efficient and effective at memorizing frequent interactions under strict latency constraints. However, they are inherently agnostic to content-level semantics, limiting their ability to leverage rich multimodal information in videos~\cite{AMMF,LEMUR}. Furthermore, they generalize poorly to long-tailed items with sparse signals~\cite{ASR,ILT}. A natural direction is to incorporate multimodal semantic embeddings to enrich item representations. Prior work such as SimTier~\cite{SimTier}, MOON~\cite{MOON}, MUSE~\cite{MUSE}, and DMF~\cite{DMF} demonstrates that dense semantic features can partially compensate for sparse interaction signals. However, such approaches often suffer from limited task adaptivity and non-trivial serving overhead. Moreover, semantic features are typically introduced as auxiliary inputs rather than being tightly integrated into ID-based ranking architectures.
    
    To address the serving overhead and limited task adaptivity of dense semantic embeddings, recent work proposes Semantic IDs (SIDs)~\cite{TIGER,SPM,QARM}, which discretize semantic representations into compact codes. Beyond efficiency, SIDs exhibit two properties desirable for large-scale ranking: they preserve hierarchical semantic structure through structured quantization, enabling coarse-to-fine abstraction within ID-based architectures, and they can be directly optimized as trainable identifiers under discriminative ranking objectives, allowing semantic representations to adapt to downstream supervision. However, existing approaches that incrementally introduce SIDs into ranking models typically treat them as static item-side features. They do not explicitly exploit the hierarchical structure of SIDs to model coarse-to-fine semantic abstraction. More importantly, they lack mechanisms to adaptively balance semantic generalization and ID-based memorization according to item popularity (head vs. long-tail). As a result, the potential of Semantic IDs to systematically improve ranking—especially for long-tailed content—remains underexplored. 

    These observations suggest that the key challenge is not merely introducing semantic representations, but how to exploit the multi-level semantic granularity of SIDs and systematically coordinate semantic generalization with sparse ID-based memorization under strict industrial constraints. We argue that effective semantic modeling in ranking systems requires a unified coordination mechanism that adapts to item popularity and jointly optimizes memorization and generalization. Motivated by this perspective, we propose \textbf{SID-Coord}, a lightweight SID framework that tightly integrates semantic modeling into ID-based ranking. SID-Coord operationalizes coordination along three dimensions. First, it performs granularity coordination via adaptive fusion of hierarchical SIDs for head–tail heterogeneity. Second, it performs signal coordination via target-aware HID–SID gating to balance memorization and generalization. Third, it performs interest coordination via distribution-level semantic alignment between candidates and histories.

\section{Related Work}
    Semantic IDs aim to discretize dense semantic representations into compact codes suitable for large-scale retrieval and ranking. TIGER~\cite{TIGER} applies residual-quantized variational autoencoders (RQ-VAE) to hierarchically quantize multimodal content embeddings, demonstrating that semantic structure can be preserved under discrete coding. Building on RQ-VAE, recent advancements also explore variable-length semantic ID strategies to adaptively adjust code depth and improve representation capacity~\cite{Kuai}. To further enhance expressiveness, OneSearch~\cite{OneSearch} combines residual quantization with optimized product quantization (RQ-OPQ), integrating learnable rotations into the quantization process. In parallel, VQ-VAE~\cite{vq-vae} learns discrete codebooks with a commitment objective, enabling end-to-end training of discrete latent representations that are also widely used for semantic discretization.

    Industrial ranking systems still predominantly rely on discriminative models trained on sparse IDs due to their robustness and serving efficiency, motivating efforts to integrate Semantic IDs into ID-based rankers. SPM~\cite{SPM} proposes compositional encoding of hierarchical semantic IDs using SentencePiece Models, achieving efficient representation under a fixed codebook size and empirically demonstrating the advantage of semantic IDs over randomly hashed IDs. Prefix-based~\cite{Prefix} methods represent semantic IDs with prefix n-grams and extend them to user behavior sequence modeling, showing improved prediction accuracy and stability in large-scale industrial systems. However, these approaches typically incorporate Semantic IDs as static features or loosely coupled auxiliary signals, without exploiting hierarchical granularity or adaptively coordinating semantic generalization with ID-based memorization. In contrast, our work treats SID integration not as a feature engineering problem, but as an explicit coordination problem between semantic abstraction and sparse-ID memorization within discriminative ranking models.

\section{Methodology}

    \begin{figure*}[t]
        \centering
        \includegraphics[width=\textwidth, height=6.3cm]{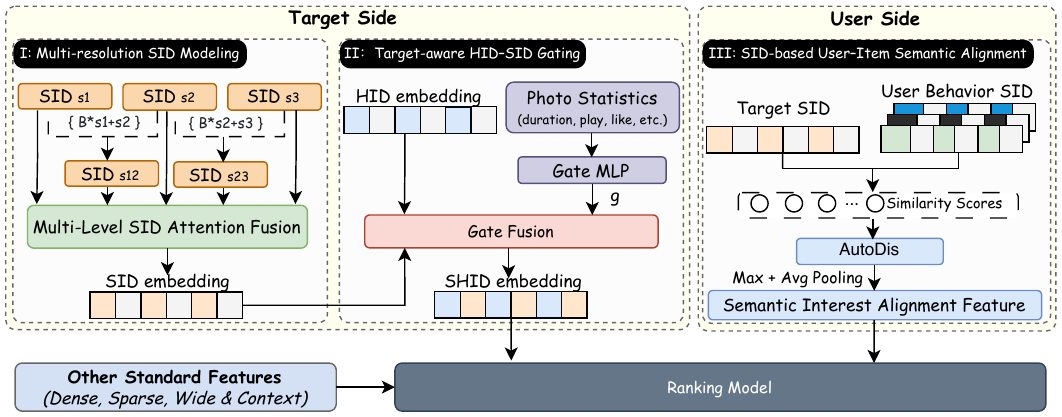}
        \caption{Overview of the proposed SID-Coord framework.}
        \label{fig:framework}
    \end{figure*}
    
    As illustrated in Figure~1, \textbf{SID-Coord} consists of three components: (1) Multi-resolution SID modeling, which adaptively fuses hierarchical SIDs to capture coarse-to-fine semantics; (2) Target-aware HID--SID gating, which balances memorization and semantic generalization; and (3) SID-based interest alignment, which models semantic consistency between the target item and the user's behavior history.

    \subsection{Multi-resolution SID Modeling}
        We model each item with multi-resolution Semantic IDs and coordinate coarse-to-fine semantics through two steps: (i) hierarchical SID composition to explicitly encode semantic refinement, and (ii) attention-based fusion to adaptively select the most effective semantic resolution for ranking.

        \textbf{Hierarchical SID composition.} For each item $v$, the content-side quantizer produces three Semantic IDs $\{s^{(1)}_v, s^{(2)}_v, s^{(3)}_v\}$ with increasing semantic specificity, following the residual-quantization (RQ) framework in QARM~\cite{QARM}. Each $s^{(i)}_v$ is a discrete integer index from an 8192-sized codebook. To explicitly encode refinement relationships while preserving hierarchical locality, we construct composite Semantic IDs between adjacent levels:
        \begin{equation}\label{eq:composition}
            s^{(1,2)}_v = B \cdot s^{(1)}_v + s^{(2)}_v, \quad s^{(2,3)}_v = B \cdot s^{(2)}_v + s^{(3)}_v, 
        \end{equation}
        where $B$ (set to 10000 in our implementation) is a base larger than the codebook size to guarantee a unique mapping from an ordered pair to a single integer. For example, with $s^{(1)}_v=17$ and $s^{(2)}_v=5301$, we obtain $s^{(1,2)}_v = 175301$. Rather than treating SIDs as a flat sequence like standard N-gram methods—which often suffer from exponential parameter explosion—this radix-$B$ construction explicitly exploits the tree-like dependency inherent in the residual quantization hierarchy. This structured composition provides two key advantages: (1) it refines the semantic space within each coarse parent region, enabling finer discrimination without breaking structural locality, and (2) it strictly prevents the indiscriminate mixing of distant scales (e.g., non-adjacent $s^{(1,3)}_v$) that would bypass intermediate semantic refinement and lead to over-fragmented, sparse representations. Furthermore, we intentionally omit the full-depth combination $s^{(1,2,3)}$, as it drastically inflates the parameter space and exacerbates data sparsity while yielding only marginal performance gains.

        \textbf{Attention-based resolution fusion.} After composition, each item is represented by five Semantic IDs $\{$ {${s}^{(1)}$, ${s}^{(2)}$, ${s}^{(3)}$, ${s}^{(1,2)}$, ${s}^{(2,3)}$ $\}$, which are mapped to dense embeddings $\{$ ${e}^{(1)}_{v}$, ${e}^{(2)}_{v}$, ${e}^{(3)}_{v}$, ${e}^{(1,2)}_{v}$, ${e}^{(2,3)}_{v}$ $\}$ via a shared embedding table. Since different items benefit from different semantic resolutions, we introduce an attention-based resolution selector (Figure~2) that learns weights over all resolutions and computes the final semantic representation as a weighted fusion. This design adaptively emphasizes the appropriate semantic granularity during ranking.

        \begin{figure}[t]
            \centering
            \includegraphics[width=\columnwidth,height=2.9cm]{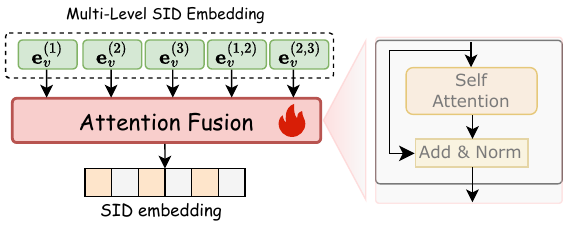} 
            \caption{SID Multi-Level Attention Fusion Mechanism.}
            \vspace{-2mm}
            \Description{..}
            \label{fig:framework}
        \end{figure}

    \subsection{Target-aware HID–SID Gating}
        HID and SID representations exhibit complementary properties: HIDs capture memorization from large-scale sparse co-occurrence signals, whereas SIDs provide semantic-level generalization. To explicitly coordinate these two signals under heterogeneous item popularity, we introduce a target-aware gating mechanism.
        
        Let $e^{hid}_v$ denote the HID embedding of the target item and $e^{sid}_v$ denote its SID representation. We use historical statistical features of the target item, such as counts of exposures, clicks, likes, shares, and comments, to predict a gating weight $g \in [0,1]$ via a lightweight gating network, which is implemented as a two-layer Multi-Layer Perceptron (MLP) with a sigmoid activation function. These statistical signals serve as a direct proxy for item popularity, enabling the network to explicitly distinguish data-sparse long-tail items from well-exposed head items, thereby learning to adaptively balance SID-based generalization and HID-based memorization. The fused representation is computed as:
        \begin{equation}
             {e}^{shid}_{v}=g \cdot {e}^{hid}_{v} + (1-g) \cdot{e}^{sid}_{v},
        \end{equation}
        
        As Figure 3 illustrates, the gating network successfully performs the intended popularity-aware coordination: ${g}$ increases with item popularity, indicating that the model assigns more weight to HID-based memorization for head items while relying more on SID-based semantic generalization for tail items. The fused representation ${e}^{shid}_{v}$ is subsequently used as the target item representation in the ranking model.

        \begin{figure}[t]
            \centering
            \includegraphics[width=\columnwidth, height=4.4cm]{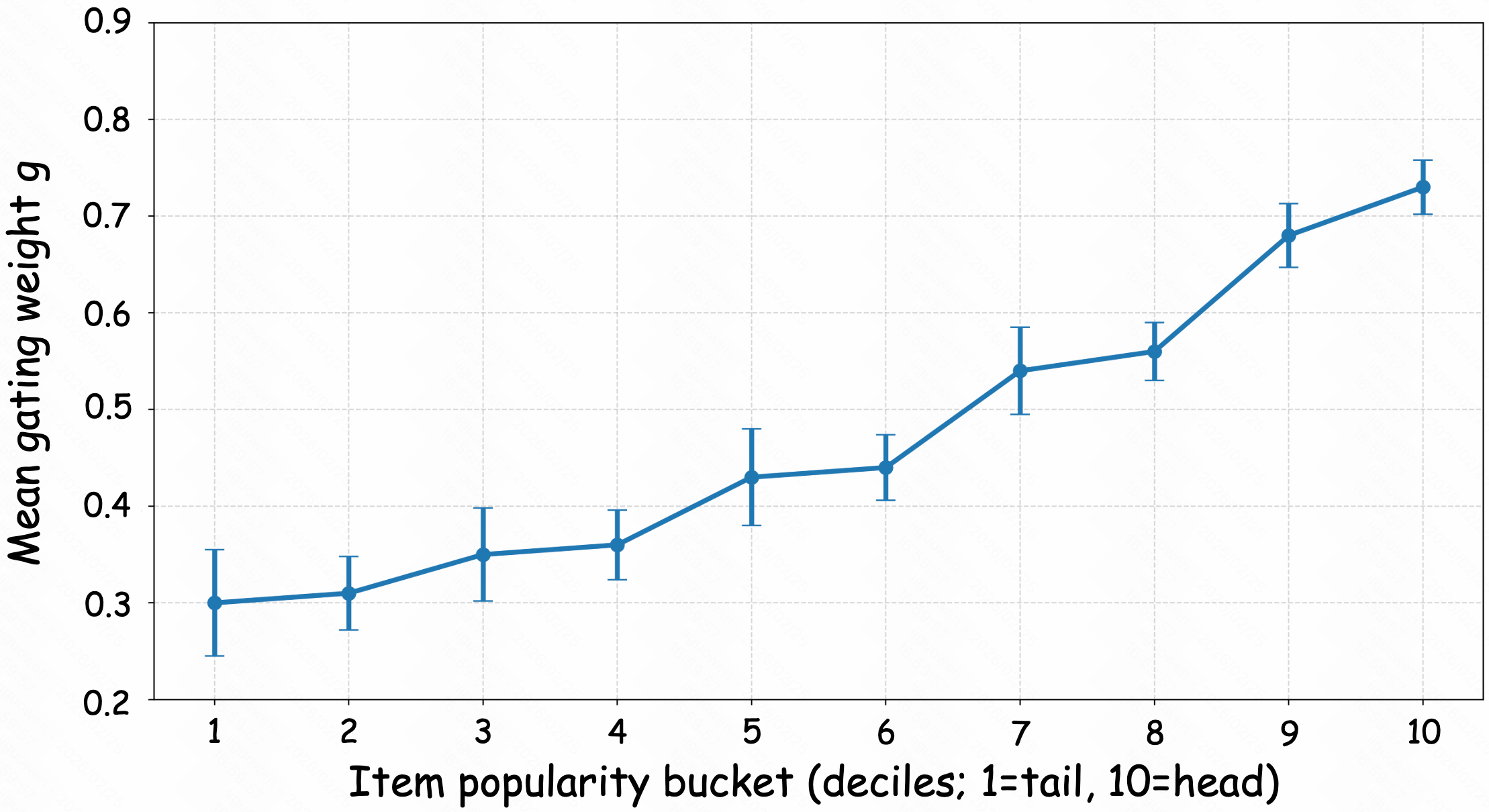}
            \caption{Analysis of Popularity-aware Coordination. Popularity buckets are defined by 7-day exposure (deciles). Points show mean item-level ${g}$, with confidence intervals.}
            \vspace{-2mm}
            \Description{..}
            \label{fig:g_popularity}
        \end{figure}

    \subsection{User–Item Semantic Alignment}
        Beyond target-side coordination, we further coordinate semantic modeling on the user side by explicitly capturing semantic alignment between the candidate item and the user’s behavior sequence. While user behavior modeling plays a pivotal role in ranking systems, traditional approaches relying solely on specific identifiers (e.g., photo or author IDs) are often semantically blind to intrinsic content correlations. We therefore incorporate SIDs to enhance content understanding and explicitly strengthen the semantic matching between historical consumption and current candidates. Given a user ${u}$ with behavior history ${H}_{u}$ = \{ ${v}_{1}$,..., ${v}_{T}$ \}, we represent each historical item ${v}_{t}$ using its semantic embedding ${e}^{sid}_{v_{t}}$ derived from Section 3.1. For a candidate item ${v}$, we compute cosine similarities between its semantic embedding ${e}^{sid}_{v}$ and the historical embeddings:
        \begin{equation}
        {S =[sim({e}^{sid}_{v},{e}^{sid}_{v_{1}}),...,sim({e}^{sid}_{v}, {e}^{sid}_{v_{T}})]},
        \end{equation}
        where $sim(\cdot)$ denotes cosine similarity. Rather than using raw pairwise similarities, we apply AutoDis~\cite{autodis} to transform the similarity scores into distribution-aware representations, capturing higher-order semantic interest patterns. We then apply max and average pooling to extract the strongest semantic match and overall consistency, respectively. Finally, these alignment features are integrated into the backbone ranking model together with other standard features to improve prediction.

\section{Experiment}

    All experiments are conducted in Kuaishou’s short-video search ranking system. Following the OneRec~\cite{QARM, OneRec} methodology, we generate foundational SIDs via RQ-KMeans with three residual codebooks of size 8192 each (i.e., 3×8192 codewords). Training data is sampled from real user interaction logs, containing approximately 4.5B examples. Evaluation is performed on a chronologically held-out set of about 130M examples collected on the following day. All methods are trained on the same data split and evaluated under identical feature sets, backbone architecture, and optimization settings, so that differences are solely attributable to how Semantic IDs are constructed and integrated.
    
    \subsection{Baselines}
        We compare SID-Coord with the production baseline and several representative SID-based methods. All models share the same ranking backbone and differ only in how Semantic IDs are composed and incorporated into the model. 
        \textbf{Prefix-Ngram} represents hierarchical Semantic IDs using prefix-based composition ~\cite{Prefix}. \textbf{Ngram} extends prefix composition to adjacent n-gram combinations ~\cite{SPM}. \textbf{SPM-SID} applies subword segmentation over SID sequences with a fixed vocabulary ~\cite{SPM}. \textbf{DAS} incorporates explicit SID matching statistics between target items and user histories ~\cite{DAS}.

        Table 1 reports the overall performance comparison. All SID-based methods outperform the production baseline, confirming the effectiveness of incorporating semantic signals into ID-based ranking. Among them, SID-Coord achieves the best performance across all metrics. In particular, SID-Coord improves AUC by +0.33\% overall and +0.42\% on long-tail items, with a substantial UAUC gain on long-tail (+0.93\% relative to baseline). It is worth noting that even fractional AUC improvements are practically meaningful in our large-scale system and translate to measurable online gains, as corroborated by our online A/B testing results in Section 4.3. These results demonstrate that our end-to-end SID-Coord framework significantly boosts long-tail performance without compromising head item accuracy.

        \begin{table}[t]
        \caption{Overall Performance Comparison.}
        \vspace{-2mm}
        \label{exp:overall}
        \centering
        \begin{tabular*}{\columnwidth}{@{\extracolsep{\fill}}lcccc@{}}
        \hline
        & \multicolumn{2}{c}{ALL} & \multicolumn{2}{c}{Long-tail} \\
        \cline{2-3}\cline{4-5}
        Model & AUC & UAUC & AUC & UAUC \\
        \hline
        base & 0.7683 & 0.6091 & 0.7691 & 0.5951 \\
        Prefix-Ngram & 0.7688 & 0.6096 & 0.7698 & 0.5962 \\
        Ngram & 0.7695 & 0.6100 & 0.7706 & 0.5961 \\
        SPM-SID & 0.7700 & 0.6101 & 0.7711 & 0.5974 \\
        DAS & 0.7693 & 0.6111 & 0.7705 & 0.5983 \\
        \textbf{SID-Coord} & 0.7708 & 0.6158 & 0.7723 & 0.6045 \\
        \hline
        \vspace{-6mm}
        \end{tabular*}
        \end{table}
        
    \subsection{Ablation Study}
        We conduct ablation studies to quantify the contribution of each component in SID-Coord. Table~2 reports results obtained by removing individual modules from the full model.

        \textbf{Effect of Multi-resolution SID Modeling (Full w/o I).}
        Removing the multi-resolution SID module causes the largest performance drop among the three components (AUC: -0.22\% overall, -0.26\% on long-tail). This suggests that adaptive semantic granularity is critical for sparse long-tail items, as it enables coarse-level statistical sharing while retaining fine-grained discrimination when sufficient signals are available.
        
        \textbf{Effect of Target-aware HID–SID Gating (Full w/o II).}
        Disabling the gating mechanism leads to consistent regressions (AUC: -0.13\% overall, -0.14\% on long-tail), with a more noticeable drop on long-tail UAUC. This indicates that simply combining HID and SID signals is insufficient; popularity-aware gating is necessary to balance HID-based memorization and SID-based generalization under heterogeneous interaction densities.

        \textbf{Effect of SID-based Interest Alignment (Full w/o III).} 
        Removing semantic alignment also degrades performance (AUC: -0.16\% overall, -0.19\% on long-tail). The consistent drop across both overall and long-tail subsets shows that modeling distribution-level semantic similarity between target items and user histories provides complementary user–item matching signals beyond item-side semantics alone.
        
       \begin{table}[t]
        \caption{Ablation study of different components.}
        \vspace{-2mm}
        \label{tab:ablation}
        \centering
        \begin{tabular*}{\columnwidth}{@{\extracolsep{\fill}}lcccc@{}}
        \hline
        & \multicolumn{2}{c}{ALL} & \multicolumn{2}{c}{Long-tail} \\
        \cline{2-3}\cline{4-5}
        Model & AUC & UAUC & AUC & UAUC \\
        \hline
        FULL          & 0.7708          & 0.6158 & 0.7723          & 0.6045 \\
        FULL w/o I    & 0.7691 (-0.22\%) & 0.6117 & 0.7703 (-0.26\%) & 0.6005 \\
        FULL w/o II   & 0.7698 (-0.13\%) & 0.6121 & 0.7712 (-0.14\%) & 0.6010 \\
        FULL w/o III  & 0.7696 (-0.16\%) & 0.6120 & 0.7708 (-0.19\%) & 0.6016 \\
        \hline
        \end{tabular*}
        \end{table}

        \begin{table}[t]
        \caption{Online A/B Testing Results.}
        \vspace{-2mm}
        \label{exp:data}
        \centering
        \begin{tabular}{lccc}
        \hline
        Model & watch time & play count & long-play rate \\
        \hline
        base & - & - & - \\
        \textbf{SID-Coord} & +0.369\% & +0.472\% & +0.664\% (+0.226 pp) \\
        \hline
        \end{tabular}
        \end{table}

    \subsection{Online A/B Testing}
        We further evaluate SID-Coord via an online A/B test in the production system. Following a 7-day A/A test to validate traffic randomization and metric stability, we run a 7-day A/B experiment with 10\% total traffic split equally between control and treatment. Statistical significance is assessed using a difference-in-differences (DiD) estimator on daily aggregated metrics, which is adopted to control for common temporal variation between groups. We report 95\% confidence intervals for all key metrics.

        As shown in Table~3, SID-Coord improves search watch time by +0.369\% (95\% CI: [0.05\%, 0.69\%]), play count by +0.472\% (95\% CI: [0.17\%, 0.77\%]), and long-play rate by +0.664\% (+0.226 pp; 95\% CI: [0.49\%, 0.84\%]). These improvements are statistically significant and remain stable throughout the experiment window.

        We also examine the serving overhead of SID-Coord in the online system. On sampled production traffic, the average search latency changes by only +0.005\% (95\% CI: [-0.08\%, 0.09\%]), indicating no statistically significant latency increase in practice. This result supports that SID-Coord can be integrated into the existing ranking system with negligible online serving cost.


            
\section{Conclusion}
    We propose SID-Coord, which explicitly coordinates multi-resolution SID-based generalization with sparse HID-based memorization for short-video search ranking. Extensive offline and online A/B tests demonstrate significant ranking improvements, yielding particularly strong gains on long-tailed items.


\bibliographystyle{ACM-Reference-Format}

\bibliography{reference}


\end{document}